\documentstyle[12pt]{article}
\begin{document}
\title{\hspace {11cm}{\small \bf IMPNWU-980316}\\
\vskip 2truecm
\bf\huge
Algebraic Bethe ansatz for the supersymmetric $t-J$ model with 
reflecting boundary conditions
}

\author{
{\bf
Heng Fan$^{a,b}$,Bo-yu Hou$^b$,Kang-jie Shi$^{a,b}$}\\
\normalsize
$^a$ CCAST(World Laboratory)\\
\normalsize
P.O.Box 8730,Beijing 100080,China\\
\normalsize
$^b$ Institute of Modern Physics, P.O.Box 105,\\
\normalsize
Northwest University, Xian,710069,China\thanks{Mailing address}\\
}

\maketitle

\begin{abstract}
In the framework of the graded quantum inverse scattering method (QISM), 
we obtain the eigenvalues and eigenvectors of the supersymmetric
$t-J$ model with reflecting boundary conditions in FFB background.
The corresponding Bethe ansatz equations are obtained.
\end{abstract}


\newpage
\renewcommand{\thesection}{\Roman{section}}

\section{Introduction}
It is believed that the strongly correlated electronic systems are
important in studying the high-$T_c$ superconductivity. An appropriate
starting point is the $t-J$ model which was proposed by
Anderson $et~al$.$^{1, 2}$ The Hamiltonian includes the 
near-neighbour hopping ($t$) and antiferromagnetic exchange ($J$).
\begin{eqnarray}
H=\sum _{j=1}^L\left\{ -t{\cal {P}}
\sum _{\sigma =\pm 1}(c_{j, \sigma}^{\dagger }
c_{j+1, \sigma }+H.c.){\cal {P}}+J({\bf S}_j{\bf S}_{j+1}
-{1\over 4}n_nn_{j+1})\right\}.
\end{eqnarray}
Essler and Korepin $et~al~$$^3$ shown that this model is supersymmetric, 
and the 
one-dimensional Hamiltonian can be obtained from the transfer matrix of 
the two-dimensional supersymmetric exactly solvable lattice model. They
used the graded QISM and obtained the eigenvalues and eigenvectors for
the supersymmetric $t-J$ model with periodic boundary
conditions in three different backgrounds. 

We know that the exactly solvable models are generally solved by imposing 
periodic boundary conditions. Recently, solvable models with reflecting
boundary conditions have been attracting a great deal of 
interests.$^{4, 5, 6, 7}$The Hamiltonian includes non-trivial
boundary terms which are determined by the boundary K matrices.
In the present paper, we will use the algebraic
Bethe ansatz method to solve the eigenvalues and eigenvectors problems of
the supersymmetric $t-J$ model with reflecting boundary conditions
in the famework of the graded QISM (FFB grading),  
and the Bethe ansatz equations are
also obtained.

The paper is organized as follows: In section II, we will introduce 
the model and the notations. In section III, we will prove the 
integrability of model with reflecting boundary conditions 
in the graded sense. The general solution of the reflection equation
is also presented in this section. In section IV, we use the algebraic
Bethe ansatz method to obtain the eigenvalues and eigenvectors of
the supersymmetric $t-J$ model. Section V includes a brief summary and
some discussions.

\section{Description of the model}
We first give a brief review of the graded version of the QISM. 
For convenient we take the natations used by Essler and Korepin.$^3$ 
And 
what we consider in this paper is the FFB grading, that is the grading 
is fermionic, fermionic and bosonic. In terms of the Grassmann parities 
this means that $\epsilon _1=\epsilon _2=1$ and $\epsilon _3=0$. The  
R is defined as 
\begin{eqnarray}
{\hat {R}}(\lambda )=b(\lambda )I+a(\lambda )\Pi ,
\end{eqnarray}
where
\begin{eqnarray}
a(\lambda )=\frac {\lambda }{\lambda +i}, 
b(\lambda )=\frac {i}{\lambda +i}.
\end{eqnarray}
And the identity operator is given by 
$I_{a_1a_2}^{b_1b_2}=\delta _{a_1b_1}\delta _{a_2b_2}$, the
matrix $\Pi $ permutes the individual linear spaces in the tensor
product space,
\begin{eqnarray}
\Pi _{a_1a_2}^{b_1b_2}=\delta _{a_1b_2}\delta _{a_2b_1}
(-1)^{\epsilon _{b_1}\epsilon _{b_2}}.
\end{eqnarray}
Explicitly, we can write the R-matrix as:
\begin{eqnarray}
{\hat {R}}(\lambda )=
\left( \begin{array}{ccccccccc}
b(\lambda )-a(\lambda )&0&0&0&0&0&0&0&0\\
0&b(\lambda )&0&-a(\lambda )&0&0&0&0&0\\       
0&0&b(\lambda )&0&0&0&a(\lambda)&0&0\\
0&-a(\lambda )&0&b(\lambda )&0&0&0&0&0\\
0&0&0&0&b(\lambda )-a(\lambda )&0&0&0&0\\
0&0&0&0&0&b(\lambda )&0&a(\lambda )&0\\
0&0&a(\lambda)&0&0&0&b(\lambda )&0&0\\
0&0&0&0&0&a(\lambda )&0&b(\lambda )&0\\
0&0&0&0&0&0&0&0&1\end{array}\right) .
\end{eqnarray}
As is well known the Yang-Baxter$^{8,9}$ relation plays a key role for
integrable models with periodic boundary conditions which takes
the form
\begin{eqnarray}
{\hat {R}}(\lambda -\mu )L_n(\lambda )\otimes L_n(\mu )
=L_n(\mu )\otimes L_n(\lambda ){\hat {R}}(\lambda -\mu ),
\end{eqnarray}
where the tensor product is in the graded sense
\begin{eqnarray}
(F\otimes G)_{ac}^{bd}=F_{ab}G_{cd}
(-1)^{\epsilon _c(\epsilon _a+\epsilon _b)}.
\end{eqnarray}
The $n$ means the $n$-th quantum space which is standard in QISM.$^{10}$ 
We can 
also write the Yang-Baxter relation explicitly as
\begin{eqnarray}
&&{\hat {R}}
(\lambda -\mu )_{a_1a_2}^{c_1c_2}L_n(\lambda )_{c_1\alpha _n}^{b_1
\gamma _n}L_n(\mu )_{c_2\gamma _n}^{b_2\beta _n}(-1)^{\epsilon _{c_2}
(\epsilon _{c_1}+\epsilon _{b_1})}
\nonumber \\
&=&L_n(\mu )_{a_1\alpha _n}^{c_1\gamma _n}
L_n(\lambda )_{a_2\gamma _n}^{c_2\beta _n}(-1)^{\epsilon _{a_2}
(\epsilon _{a_1}+\epsilon _{c_1})}
{\hat {R}}(\lambda -\mu )_{c_1c_2}^{b_1b_2}.
\end{eqnarray}
The $L$ operator can be constructed from the R-matrix
\begin{eqnarray}
L_n(\lambda )_{a\alpha }^{b\beta }
=\Pi _{a\alpha }^{c\gamma }{\hat {R}}(\lambda )_{c\gamma }^{b\beta }
=[b(\lambda )\Pi +a(\lambda )I]_{a\alpha }^{b\beta }.
\end{eqnarray}
So, the $L$ operator is of the form 
\begin{eqnarray}
L_n(\lambda )=
\left(\begin{array}{ccc}
a(\lambda )-b(\lambda )e_n^{11} &-b(\lambda )e_n^{21} & 
b(\lambda )e_n^{31}\\
-b(\lambda )e_n^{12}& a(\lambda -b(\lambda )e_n^{22} &
b(\lambda )e_n^{32}\\
b(\lambda )e_n^{13}&b(\lambda )e_n^{23}&
a(\lambda )+b(\lambda )e_n^{33}\end{array}
\right),
\end{eqnarray}
where $e_n^{ab}$ are quantum operators acting in the $n$-th quantum 
space with matrix representation $(e_n^{ab})_{\alpha \beta }=
\delta _{a\alpha }\delta _{b\beta}$. 
The monodromy matrix $T_L(\lambda )$ defined as the matrix product over
the $L$ operators on all site of the lattice, 
\begin{eqnarray}
T_L(\lambda )=L_L(\lambda )L_{L-1}(\lambda )\cdots L_1(\lambda ), 
\end{eqnarray}
where the tensor product is still in the graded sense, and we will not 
point it out in the following.
\begin{eqnarray}
&&\{ [T_L(\lambda )]^{ab}\}_{\begin {array}{c}
\alpha _1\cdots \alpha _L\\
\beta _1\cdots \beta _L\end{array}}
\nonumber \\
&=&L_L(\lambda )_{a\alpha _L}^{c_L\beta _L}
L_{L-1}(\lambda )_{c_L\alpha _{L-1}}^{c_{L-1}\beta _{L-1}}
\cdots L_1(\lambda )_{c_2\alpha _1}^{b\beta _1}
(-1)^{\sum _{j=2}^L(\epsilon _{\alpha _j}+\epsilon _{\beta _j})
\sum _{i=1}^{j-1}\epsilon _{\alpha _i}}
\end{eqnarray}
By repeatedly using the Yang-Baxter relation (6), one can prove easily that
the monodromy matrix also satisfy the Yang-Baxter relation
\begin{eqnarray}
{\hat {R}}(\lambda -\mu )[T_L(\lambda )\otimes T_L(\mu )]
=T_L(\mu )\otimes T_L(\lambda ){\hat {R}}(\lambda -\mu ),
\end{eqnarray}

The transfer matrix $\tau _{peri}(\lambda )$ 
of this model is defined as the supertrace of the
monodromy matrix in the auxiliary space. It is defined as the following 
in the general case
\begin{eqnarray}
\tau _{peri}(\lambda )=str[T_L(\lambda )]
=\sum (-1)^{\epsilon _a}[T_L(\lambda )]^{aa}.
\end{eqnarray}
For the case considered in this paper, if we represent
\begin{eqnarray}
T_L(\lambda )=
\left( \begin{array}{ccc}
A_{11}(\lambda )&A_{12}(\lambda ) &B_1(\lambda )\\
A_{21}(\lambda )&A_{22}(\lambda )&B_2(\lambda )\\
C_1(\lambda )&C_2(\lambda )&D(\lambda )\end{array}
\right).
\end{eqnarray}
The transfer matrix is then given as
\begin{eqnarray}
\tau (\lambda )_{peri}=-A_{11}(\lambda )-A_{22}(\lambda )+D(\lambda ).
\end{eqnarray}
As a consequence of the Yang-Baxter relation (13), we can prove that 
the transfer matrix commute with each other for different spectrum
parameters.
\begin{eqnarray}
[\tau _{peri}(\lambda ),\tau _{peri}(\mu )]=0 
\end{eqnarray}
It has been proved that the Hamiltonian obtained by taking the first
logarithmic derivative at zero spectral parameter 
\begin{eqnarray}
H_{(2)}=-i\frac {d\ln [\tau (\lambda )]}{d\lambda }|_{\lambda =0}
=-\sum _{k=1}^L\Pi ^{k,k+1}
\end{eqnarray}
is equivalent to the Hamiltonian of the supersymmtric $t-J$ model.$^3$
Here we have omitted the indentities.

What mentioned above is for periodic boundary conditions. We will study 
the case of the reflecting boundary conditions for the supersymmetric
$t-J$ model. For convenience, we change the braided R-matrix 
$\hat {R}$ (2,5) to the non-braided R-matrix
\begin{eqnarray}
R(\lambda )&=&b(\lambda )\Pi +a(\lambda )I\nonumber \\
&=&
\left( \begin{array}{ccccccccc}
a(\lambda )-b(\lambda )&0&0&0&0&0&0&0&0\\
0&a(\lambda )&0&-b(\lambda )&0&0&0&0&0\\       
0&0&a(\lambda )&0&0&0&b(\lambda)&0&0\\
0&-b(\lambda )&0&a(\lambda )&0&0&0&0&0\\
0&0&0&0&a(\lambda )-b(\lambda )&0&0&0&0\\
0&0&0&0&0&a(\lambda )&0&b(\lambda )&0\\
0&0&b(\lambda)&0&0&0&a(\lambda )&0&0\\
0&0&0&0&0&b(\lambda )&0&a(\lambda )&0\\
0&0&0&0&0&0&0&0&1\end{array}\right) .
\nonumber \\
\end{eqnarray}
So, we change the Yang-Baxter relation (13) as:
\begin{eqnarray}
R(\lambda -\mu )T_1(\lambda )T_2(\mu )
=T_2(\mu )T_1(\lambda )R(\lambda -\mu )
\end{eqnarray}
Here $1, 2$ mean the auxiliary space. The definition for monodromy
matrix $T$ remain the same as before.

\section{Integrability of the supersymmetric $t-J$ model with  
reflecting boundary conditions}
As we know the Yang-Baxter relation is enough to prove the
integrability of the exactly solvable model. For the reflecting
boundary conditions, besides the Yang-Baxter relation, we also
need the reflection equation and the dual reflection equation
to prove the integrability of the solvable model. The reflection
equation was first proposed by Cherednik.$^{11}$ In order to 
prove the integrability for reflecting boundary conditions, 
Sklyanin$^4$ proposed the dual reflection equation. 
Generally, the dual reflection equation which depends on
the unitarity and cross-unitrarity relations of the R-matrix   
takes different forms for different models. $^7$

For the R-matrix considered in this paper (19), one can prove that
the R-matrix satisfy the unitarity relation
\begin{eqnarray}
R_{12}(\lambda )R_{21}(-\lambda )=1,
\end{eqnarray}
where $R_{21}=\Pi R_{12}\Pi $. We can also find that this R-matrix
has a symmetry $R_{21}=R_{12}$.

We define the super-transposition $st$ as:
\begin{eqnarray}
(A^{st})_{ij}=A_{ji}(-1)^{(\epsilon _i+1)\epsilon _j}.
\end{eqnarray}
For the case considered in this paper $\epsilon _1=\epsilon _2=1
, \epsilon _3=0$, we can rewrite the above relation explicitly as:
\begin{eqnarray}
\left( \begin{array}{ccc}
A_{11}&A_{12}&B_1\\
A_{21}&A_{22}&B_2\\
C_1&C_2&D\end{array}\right)^{st}
=\left( \begin{array}{ccc}
A_{11}&A_{21}&C_1\\
A_{12}&A_{22}&C_2\\
-B_1&-B_2&D\end{array}\right)
\end{eqnarray}
Here for convenience, we also define the inverse of 
the super-transposition $\bar {st}$ as 
$\{ A^{st}\} ^{\bar {st}}=A$.

Considering the R-matrix presented above, we find that the R-matrix 
satisfy the following cross-unitarity relation
\begin{eqnarray}
R_{12}^{st_1}(i-\lambda )R_{21}^{st_1}(\lambda )=\rho (\lambda ),
\nonumber \\
\rho (\lambda )=\frac {(i-\lambda )\lambda }{(\lambda +i)(2i-\lambda )},
\end{eqnarray}
here $st_1$ means taking super-transposition in the first space.

Next, we introduce the graded version of the reflection equation as:
\begin{eqnarray}
&&R_{12}(\lambda -\mu )K_1(\lambda )R_{21}(\lambda +\mu )
K_2(\mu )
\nonumber \\
&=&K_2(\mu )R_{12}(\lambda +\mu )K_1(\lambda )R_{21}(\lambda -\mu ).
\end{eqnarray}
It can also be rewritten as:
\begin{eqnarray}
&&R(\lambda -\mu )_{a_1a_2}^{b_1b_2}
K(\lambda )_{b_1}^{c_1}R(\lambda +\mu )_{b_2c_1}^{c_2d_1}
K(\mu )_{c_2}^{d_2}
(-1)^{(\epsilon _{b_1}+\epsilon _{c_1})\epsilon _{b_2}}
\nonumber \\
&=&K(\mu )_{a_2}^{b_2}
R(\lambda +\mu )_{a_1b_2}^{b_1c_2}
K(\lambda )_{b_1}^{c_1}R(\lambda -\mu )_{c_2c_1}^{d_2d_1}
(-1)^{(\epsilon _{b_1}+\epsilon _{c_1})\epsilon _{c_2}}.
\end{eqnarray}
Here the refleting $K$ is the solution of the reflection equation. 
We will just consider the diagonal $K$ matrix in this paper, so 
we suppose 
\begin{eqnarray}
K(\lambda )_a^b=\delta _{ab}k_a(\lambda ). 
\end{eqnarray}
Subsititing this condition into the reflection equation (26), we find
the only non-trivial relation is 
\begin{eqnarray}
&&R(\lambda -\mu )_{a_1a_2}^{a_1a_2}R(\lambda +\mu )_{a_2a_1}^{a_1a_2}
k(\lambda )_{a_1}k(\mu )_{a_1}
+R(\lambda -\mu )_{a_1a_2}^{a_2a_1}R(\lambda +\mu )_{a_1a_2}^{a_1a_2}
k(\lambda )_{a_2}k(\mu )_{a_1}
\nonumber \\
&=&R(\lambda +\mu )_{a_1a_2}^{a_1a_2}R(\lambda -\mu )_{a_2a_1}^{a_1a_2}
k(\mu )_{a_2}k(\lambda )_{a_1}
+R(\lambda +\mu )_{a_1a_2}^{a_2a_1}
R(\lambda -\mu )_{a_1a_2}^{a_2a_1}k(\mu )_{a_2}k(\lambda )_{a_2}.
\nonumber \\
\end{eqnarray}
Solving this relation, we can find two different types of solution
to the graded reflection equation
\begin{eqnarray}
K_I(\lambda )
=\left( \begin{array}{ccc}\xi +\lambda &&\\
&\xi +\lambda &\\
&&\xi -\lambda \end{array}\right),  \\
K_{II}(\lambda )
=\left( \begin{array}{ccc}\xi +\lambda &&\\
&\xi -\lambda &\\
&&\xi -\lambda \end{array}\right),
\end{eqnarray}
where $\xi $ is an arbitrary parameter. According to the form
of the cross-unitarity relation of the R-matrix (24), we propose
the graded dual reflection equation take the form
\begin{eqnarray}
&&R_{12}(\mu -\lambda )K_1^+(\lambda )R_{21}(i-\lambda -\mu )
K_2^+(\mu )
\nonumber \\
&=&K_2^+(\mu )R_{12}(i-\lambda -\mu )K_1^+(\lambda )R_{21}(\mu -\lambda ).
\end{eqnarray}
One can find that there is an isomorphism between the reflection 
equation (25) and dual reflection equation (31): Given a solution of
the reflection equation (25), we can also find a solution of the 
dual reflection equation (31). But in the sense of the commuting
transfer matrix, the reflection equation and the dual reflection equation
are independent of each other. We have two types of the solutions of 
the dual reflection equation
\begin{eqnarray}
K_I^+(\lambda )
=\left( \begin{array}{ccc}\xi ^+-\lambda &&\\
&\xi ^+-\lambda &\\
&&\xi ^+-i+\lambda \end{array}\right),  \\
K_{II}^+(\lambda )
=\left( \begin{array}{ccc}\xi ^+-\lambda &&\\
&\xi ^+-i+\lambda &\\
&&\xi ^+-i+\lambda \end{array}\right),
\end{eqnarray}
here $\xi ^+$ is an arbitrary parameter which is independent of $\xi $.

Note here that there are two different types of solutions $K$ and $K^+$,
respectively, 
and we know that $K$ and $K^+$ are independent of each other in the sense of
the transfer matrix, so there are four different transfer matrix altogether
corresponding to those $K$ and $K^+$: 
$\{ K^+_I, K_I\}$; $\{ K^+_I, K_{II}\}$; $\{ K^+_{II}, K_I\}$;
$\{ K^+_{II}, K_{II}\}$.

Following the method of Sklyanin,$^4$ we define the
double-row monodromy matrix for the case of reflecting boundary 
conditions
\begin{eqnarray}
{\cal {T}}(\lambda )=T(\lambda )K(\lambda )T^{-1}(-\lambda ).
\end{eqnarray}
Using the Yang-Baxter relation (20), one can prove easily that
this double-row monodromy matrix also satisfy the reflection equation (25)
\begin{eqnarray}
&&R_{12}(\lambda -\mu ){\cal {T}}_1(\lambda )R_{21}(\lambda +\mu )
{\cal {T}}_2(\mu )
\nonumber \\
&=&{\cal {T}}_2(\mu )R_{12}(\lambda +\mu )
{\cal {T}}_1(\lambda )R_{21}(\lambda -\mu ).
\end{eqnarray}
We thus define the
tranfer matrix with open boundary conditions as:
\begin{eqnarray}
t(\lambda )=strK^+(\lambda )\cal {T}(\lambda ),
\end{eqnarray}
As before $str$ means super trace.
Next, we will prove that the defined transfer matrices with
different spectral parameters commute with each other. 
We generally in this sense mean the model is integrable.

We first take super transposition in the first space.
\begin{eqnarray}
t(\lambda )t(\mu )&=&str_1K_1^+(\lambda ){\cal {T}}_1(\lambda )
str_2K_2^+(\mu ){\cal {T}}_2(\mu )
\nonumber \\
&=&str_{12}K_1^+(\lambda )^{st_1}K_2^+(\mu ){\cal {T}}_1^{st_1}(\lambda )
{\cal {T}}_2(\mu ) ,\nonumber 
\end{eqnarray}
Now we insert the
cross-unitrarity
relation (24) of the R-matrix, and take inverse of
super transposition in the
first space, we have
\begin{eqnarray}
\cdots &=&\frac {1}{\rho (\lambda +\mu )}
str_{12}K_1^+(\lambda )^{st_1}K_2^+(\mu )
R_{12}^{st_1}(i-\lambda -\mu )R_{21}^{st_1}(\lambda +\mu )
{\cal {T}}_1^{st_1}(\lambda )
{\cal {T}}_2(\mu ) ,\nonumber \\
&=&\frac {1}{\rho (\lambda +\mu )}
str_{12}\{ K_1^+(\lambda )^{st_1}K_2^+(\mu )
R_{12}^{st_1}(i-\lambda -\mu )\} ^{\bar {st_1}}
\nonumber \\
&&~~~\{ {\cal {T}}_1(\lambda )R_{21}(\lambda +\mu )
{\cal {T}}_2(\mu ) \},\nonumber \\
&=&\frac {1}{\rho (\lambda +\mu )}
str_{12}\{ K_2^+(\mu )
R_{12}(i-\lambda -\mu )K_1^+(\lambda )
\} \{ {\cal {T}}_1(\lambda )R_{21}(\lambda +\mu )
{\cal {T}}_2(\mu ) \} .
\nonumber 
\end{eqnarray}
Insert the untarity relation of the R-matrix (21), and use the RE (35) 
and the dual RE (31), we have
\begin{eqnarray}
\cdots &=&\frac {1}{\rho (\lambda +\mu )}
str_{12}\{ K_2^+(\mu )
R_{12}(i-\lambda -\mu )K_1^+(\lambda )
R_{21}(\mu -\lambda )\} \nonumber \\
&&~~~\{ R_{12}(\lambda -\mu ){\cal {T}}_1(\lambda )R_{21}(\lambda +\mu )
{\cal {T}}_2(\mu ) \} .
\nonumber \\
&=&\frac {1}{\rho (\lambda +\mu )}
str_{12}\{ R_{12}(\mu -\lambda )K_1^+(\lambda )
R_{21}(i-\mu -\lambda )K_2^+(\mu )
\} \nonumber \\
&&~~~\{ {\cal {T}}_2(\mu )
R_{12}(\lambda +\mu ){\cal {T}}_1(\lambda )R_{21}(\lambda -\mu )
 \} .
\end{eqnarray}
Applying almost the same procedure as before,
use again the unitarity relation (21) and
the cross-unitarity relation (24), we have
\begin{eqnarray}
\cdots &=&\frac {1}{\rho (\lambda +\mu )}
str_{12}\{ K_1^+(\lambda )
R_{21}(i-\mu -\lambda )K_2^+(\mu )\} 
\{ {\cal {T}}_2(\mu )
R_{12}(\lambda +\mu ){\cal {T}}_1(\lambda )\} .
\nonumber \\
&=&\frac {1}{\rho (\lambda +\mu )}
str_{12}\{ R_{21}^{st_1}(i-\mu -\lambda )K_1^+(\lambda )^{st_1}
K_2^+(\mu )\} 
\nonumber \\
&&~~~\{ {\cal {T}}_2(\mu ){\cal {T}}_1^{st_1}(\lambda )
R_{12}^{st_1}(\lambda +\mu )\} .
\nonumber \\
&=&str_2K_2^+(\mu ){\cal {T}}_2(\mu )
str_1K_1^+(\lambda ){\cal {T}}_1(\lambda )
\nonumber \\
&=&t(\mu )t(\lambda ).
\end{eqnarray}
Thus we have proved that the transfer matrix constitute a commuting
family which gives an infinite set of conserved quantities.

Corresponding to this transfer matrix, we can also obtain the 
Hamiltonian:
\begin{eqnarray}
H_{(2)}^{Bound.}
=-{i\over 2}\frac {d\ln [t(\lambda )]}{d\lambda }|_{\lambda =0}
=-\sum _{k=1}^{L-1}\Pi ^{k,k+1}-{i\over 2}K'_1(0)
-\frac {str_1K^+_1(0)\Pi ^{L,1}}{strK^+(0)}.
\nonumber 
\end{eqnarray}
The boundary terms are determined by the reflecting K matrices.

\section{Algebraic Bethe ansatz method}
\subsection{Transfer matrix and the vacuum state}  
According to the definition of the monodromy matrix (11), we can write
the inverse of the monodromy matrix as:
\begin{eqnarray}
T^{-1}(-\lambda )
&=&L_1^{-1}(-\lambda )L_2^{-1}(-\lambda )\cdots L_L^{-1}(-\lambda )
\end{eqnarray}
With the help of the 
definition relation (34), we can rewrite the double-row monodromy matrix 
explicitly as:
\begin{eqnarray}
{\cal {T}}(\lambda )
&=&T(\lambda )K(\lambda )T^{-1}(-\lambda )
\nonumber \\&=&
\left( \begin{array}{ccc}
A_{11}(\lambda )&A_{12}(\lambda ) &B_1(\lambda )\\
A_{21}(\lambda )&A_{22}(\lambda )&B_2(\lambda )\\
C_1(\lambda )&C_2(\lambda )&D(\lambda )\end{array}
\right)\times 
\left( \begin{array}{ccc}
k_1(\lambda )&0&0\\
0&k_2(\lambda )&0\\
0&0&k_3(\lambda )\end{array}
\right)
\nonumber \\
&\times&
\left( \begin{array}{ccc}
\bar {A}_{11}(-\lambda )&\bar {A}_{12}(-\lambda ) &\bar {B}_1(-\lambda )\\
\bar {A}_{21}(-\lambda )&\bar {A}_{22}(-\lambda )&
\bar {B}_2(-\lambda )\\
\bar {C}_1(-\lambda )&\bar {C}_2(-\lambda )&\bar {D}(-\lambda )\end{array}
\right)
\nonumber \\
&=&\left( \begin{array}{ccc}
{\cal {A}}_{11}(\lambda )&{\cal {A}}_{12}(\lambda ) &
{\cal {B}}_1(\lambda )\\
{\cal {A}}_{21}(\lambda )&{\cal {A}}_{22}(\lambda )
&{\cal {B}}_2(\lambda )\\
{\cal {C}}_1(\lambda )&{\cal {C}}_2(\lambda )&{\cal {D}}
(\lambda )\end{array}
\right) .
\end{eqnarray}
For the periodic boundary conditions, Essler and Korepin $^3$choose the 
reference state in the $k$-th quantum space and the vacuum $|0>$ as: 
\begin{eqnarray}
|0>_n=\left(\begin{array}{c}
0\\0\\1\end{array}\right), 
|0>=\otimes _{k=1}^L|0>_k.
\end{eqnarray}
What we study in this paper is the
the case of the reflecting boundary conditions, we assume the vacuum state
remain the same as the case of periodic boundary conditions. 
That means the above state $|0>$ is still the vacuum state
for the reflecting boundary conditions. According to the 
definition of the monodromy matrix $T(\lambda )$ and the inverse of the
monodromy matrix $T^{-1}(\lambda )$, we have the following results:
\begin{eqnarray}
T(\lambda )|0>
&=&
\left( \begin{array}{ccc}
[a(\lambda )]^L&0 &0\\
0&[a(\lambda )]^L&0\\
C_1(\lambda )&C_2(\lambda )&1\end{array}
\right)|0>
\nonumber \\
T^{-1}(-\lambda )|0>&=&
\left( \begin{array}{ccc}
[a(\lambda )]^L&0 &0\\
0&[a(\lambda )]^L&0\\
\bar {C}_1(-\lambda )&\bar {C}_2(-\lambda )&1\end{array}
\right)|0>.
\end{eqnarray}

Now let us see the values of the double-row monodromy matrix 
${\cal {T}}$ acting on the vacuum state. One can obtain easily 
\begin{eqnarray}
{\cal {D}}(\lambda )|0>=k_3(\lambda )D(\lambda ){\bar {D}}(-\lambda )|0>
=k_3(\lambda )|0>, 
\\
{\cal {B}}_1(\lambda )|0>=0, 
~~~{\cal {B}}_2(\lambda )|0>=0,
\\
{\cal {C}}_1(\lambda )|0>\not= 0, 
~~~{\cal {C}}_2(\lambda )|0>\not= 0.
\end{eqnarray}
It is non-trivial for the other elements,  
\begin{eqnarray}
{\cal {A}}_{12}(\lambda )|0>&=&k_3(\lambda )B_1(\lambda )
{\bar {C}}_2(-\lambda )|0>, 
\nonumber \\
{\cal {A}}_{21}(\lambda )|0>&=&k_3(\lambda )B_2(\lambda )
{\bar {C}}_1(-\lambda )|0>,
\\
{\cal {A}}_{22}(\lambda )|0>&=&
[k_2(\lambda )A_{22}(\lambda ){\bar {A}}_{22}(-\lambda )+
k_3(\lambda )B_2(\lambda )
{\bar {C}}_2(-\lambda )]|0>, 
\\
{\cal {A}}_{11}(\lambda )|0>&=&
[k_1(\lambda )A_{11}(\lambda ){\bar {A}}_{11}(-\lambda )+
k_3(\lambda )B_1(\lambda )
{\bar {C}}_1(-\lambda )]|0>.
\end{eqnarray}
In order to obtain the results of the above relations, we should  
use the graded Yang-Baxter relation. From relation (20), we
can find the following explicit relation
\begin{eqnarray}
&&[T^{-1}(-\lambda )]_{a_2}^{b_2}
R(2\lambda )_{a_1b_2}^{b_1c_2}T(\lambda )_{b_1}^{c_1}
(-1)^{(\epsilon _{b_1}+\epsilon _{c_1})\epsilon _{c_2}}
\nonumber \\
&=&T(\lambda )_{a_1}^{b_1}R(2\lambda )_{b_1a_2}^{c_1b_2}
[T^{-1}(-\lambda )]_{b_2}^{c_2}(-1)^{(\epsilon _{a_1}
+\epsilon _{b_1})\epsilon _{a_2}}.
\end{eqnarray}
Acting the two sides of this relation on the vacuum state, and taking
special values for the indecies, for cases $a_1=1, a_2=3, c_1=3, c_2=2$ 
and $a_1=2, a_2=3, c_1=3, c_2=1$, with the help of relation (42), we have
the results:
\begin{eqnarray}
{\cal {A}}_{12}(\lambda )|0>&=&0, 
\nonumber \\
{\cal {A}}_{21}(\lambda )|0>&=&0.
\end{eqnarray}
For case $a_1=2, a_2=3, c_1=3, c_2=2$, we have
\begin{eqnarray}
B_2(\lambda ){\bar {C}}_2(-\lambda )|0>=
b(2\lambda ){\bar {D}}(-\lambda )D(\lambda )|0>-
b(2\lambda )A_{22}(\lambda ){\bar {A}}_{22}(-\lambda )|0>.
\end{eqnarray}
Substituting this relation into relation (47), we find
\begin{eqnarray}
{\cal {A}}_{22}(\lambda )|0>
=\{ [k_2(\lambda )-k_3(\lambda )b(2\lambda )]a^{2L}(\lambda )
+b(2\lambda )k_3(\lambda )\} |0>,
\end{eqnarray}
here for convenience, we introduce a transformation
\begin{eqnarray}
{\cal {A}}_{22}(\lambda )=\tilde {\cal {A}}_{22}(\lambda )
+b(2\lambda ){\cal {D}}(\lambda ).
\end{eqnarray}
Thus we can find the value of the element 
$\tilde {\cal {A}}_{22}$ acting on vacuum state
\begin{eqnarray}
\tilde {\cal {A}}_{22}(\lambda )|0>=
[k_2(\lambda )-k_3(\lambda )b(2\lambda )]a^{2L}(\lambda )|0>.
\end{eqnarray}
The above transformation is very important in the later 
algebraic Bethe ansatz method. Instead of  
${\cal {A}}_{22}(\lambda )$, we use $\tilde {\cal {A}}_{22}(\lambda )$
acting on the assumed eigenvectors, so we find that there are only
one wanted term which is necessary for the algebraic Bethe anstaz method.
Similarly, we have the relation
\begin{eqnarray}
B_1(\lambda ){\bar {C}}_1(-\lambda )|0>=
b(2\lambda ){\bar {D}}(-\lambda )D(\lambda )|0>-
b(2\lambda )A_{11}(\lambda ){\bar {A}}_{11}(-\lambda )|0>.
\end{eqnarray}
Introduce similar transformation
\begin{eqnarray}
{\cal {A}}_{11}(\lambda )=\tilde {\cal {A}}_{11}(\lambda )
+b(2\lambda ){\cal {D}}(\lambda ).
\end{eqnarray}
We have
\begin{eqnarray}
\tilde {\cal {A}}_{11}(\lambda )|0>=
[k_1(\lambda )-k_3(\lambda )b(2\lambda )]a^{2L}(\lambda )|0>.
\end{eqnarray}
We summarize the above results
\begin{eqnarray}
\tilde {\cal {A}}_{11}(\lambda )|0>&=&
W_1(\lambda )a^{2L}(\lambda )|0>,
\nonumber \\
\tilde {\cal {A}}_{22}(\lambda )|0>&=&
W_2(\lambda )a^{2L}(\lambda )|0>,
\nonumber \\
{\cal {D}}(\lambda )|0>&=&U^I_3(\lambda )|0>
\end{eqnarray}
Corresponding to two different types of solutions $K$ 
of the reflection equation. 
$W_j(\lambda ), j=1,2,$ and $U_3(\lambda )$ take following 
values

For $K_I(\lambda )$:
\begin{eqnarray}
W_1(\lambda )
&=&\frac {2\lambda (\lambda +\xi +i)}{2\lambda +i}
\nonumber \\
W_2(\lambda )
&=&\frac {2\lambda (\lambda +\xi +i)}{2\lambda +i}
\nonumber \\
U_3(\lambda )&=&(\xi -\lambda )
\end{eqnarray}
\indent For $K_{II}(\lambda )$:
\begin{eqnarray}
W_1(\lambda )
&=&\frac {2\lambda (\lambda +\xi +i)}{2\lambda +i}
\nonumber \\
W_2(\lambda )
&=&\frac {2\lambda (\xi -\lambda )}{2\lambda +i}
\nonumber \\
U_3(\lambda )&=&(\xi -\lambda )
\end{eqnarray}

Considering the transformation (53,56) and definition of the transfer 
matrix with reflecting boundary
conditions, we can rewrite the transfer matrix as:
\begin{eqnarray}
t(\lambda )&=&strK^+(\lambda ){\cal {T}}(\lambda )
=-k_1^+(\lambda ){\cal {A}}_{11}(\lambda )
-k_2^+(\lambda ){\cal {A}}_{22}(\lambda )
+k_3^+(\lambda ){\cal {D}}(\lambda )
\nonumber \\
&=&-W_1^+(\lambda )\tilde {\cal {A}}_{11}(\lambda )
-W_2^+(\lambda )\tilde {\cal {A}}_{22}(\lambda )
+U_3^+(\lambda ){\cal {D}}(\lambda )
\end{eqnarray}
Here $W_j^+,j=1,2$ and $U^+_3$ take the following form 

For $K_I^+(\lambda )$:
\begin{eqnarray}
W_1^+(\lambda )&=&\xi ^+-\lambda , 
\nonumber \\
W_2^+(\lambda )&=&\xi ^+-\lambda ,
\nonumber \\
U_3^+(\lambda )&=&\frac {(2\lambda -i)(\xi ^++\lambda +i)}{2\lambda +i},
\end{eqnarray}
\indent For $K_{II}^+(\lambda )$:
\begin{eqnarray}
W_1^+(\lambda )&=&\xi ^+-\lambda , 
\nonumber \\
W_2^+(\lambda )&=&\xi ^+-i+\lambda ,
\nonumber \\
U_3^+(\lambda )&=&\frac {(2\lambda -i)(\xi ^++\lambda )}{2\lambda +i}.
\end{eqnarray}

\subsection{Commutation relations and the first step of the
nested algebraic Bethe ansatz method} 
For the algebraic Bethe ansatz method, we should obtain the commutation
relations between the elements of ${\cal {T}}$. In the case of reflecting
boundary condition, instead of the Yang-Baxter relation, we need the 
reflection equation (35) to obtain the necessary commutation relations.
First of all, we introduce a transformation
\begin{eqnarray}
{\cal {A}}_{ab}(\lambda )=\tilde {\cal {A}}_{ab}(\lambda )+
\delta _{ab}b(2\lambda ){\cal {D}}(\lambda )
\end{eqnarray}
which is consistent with the former transformations (53, 56). Next We 
intend to find the commuation relations 
between $\tilde {\cal {A}}_{aa},~{\cal {D}}$ and ${\cal {C}}_b$.
The commuation relation between 
${\cal {A}}_{aa}$ and ${\cal {C}}_b$ is not necessary, because there
will appear two or more wanted terms which can not be handle for
the algebraic Bethe ansatz method. For convenience, we write the
reflection equation explicitly as:
\begin{eqnarray}
R(\lambda -\mu )_{a_1a_2}^{b_1b_2}{\cal {T}}(\lambda )_{b_1}^{c_1}
R_{21}(\lambda +\mu )_{c_1b_2}^{d_1c_2}{\cal {T}}(\mu )_{c_2}^{d_2}
(-1)^{(\epsilon _{b_1}+\epsilon _{c_1})\epsilon _{b_2}}
\nonumber \\
={\cal {T}}(\mu )_{a_2}^{b_2}
R(\lambda +\mu )_{a_1b_2}^{b_1c_2}{\cal {T}}(\lambda )_{b_1}^{c_1}
R_{21}(\lambda -\mu )_{c_1c_2}^{d_1d_2}
(-1)^{(\epsilon _{b_1}+\epsilon _{c_1})\epsilon _{c_2}}
\end{eqnarray}
Take special values for the indecies of this reflection equation,
for case $a_1=a_2=3, d_1, d_2\not= 3$, we find
\begin{eqnarray}
{\cal {C}}_{d_1}(\lambda ){\cal {C}}_{d_2}(\mu )
=-{\cal {C}}_{c_2}(\mu ){\cal {C}}_{c_1}(\lambda )
R(\lambda -\mu )_{c_2c_1}^{d_2d_1}.
\end{eqnarray}
That means that ${\cal {C}}(\lambda )$ and ${\cal {C}}(\mu )$ are  
commutative up to a scalar. Later we will use this property to construct
the eigenvector of the transfer matrix. Note here all indices in the
commutation relation take values $1, 2$. For other commutation relations,
this is also true and we will not point it out later.
For case $a_1=a_2=d_2=3, d_1\not= 3$, and considering the 
transformation (61), we have the commutation relation between
${\cal {D}}$ and ${\cal {C}}$,
\begin{eqnarray}
{\cal {D}}(\lambda ){\cal {C}}_{d}(\mu )
&=&\frac {(\lambda +\mu )(\lambda -\mu -i)}{(\lambda +\mu +i)
(\lambda -\mu )}{\cal {C}}_d(\mu ){\cal {D}}(\lambda )
\nonumber \\
&+&\frac {2i\mu }{(\lambda -\mu )(2\mu +i)}{\cal {C}}_d(\lambda )
{\cal {D}}(\mu )
-\frac {i}{\lambda +\mu +i}{\cal {C}}_b(\lambda )
\tilde {\cal {A}}_{bd}(\mu ).
\end{eqnarray}

To obtain the  commutation relation between $\tilde {\cal {A}}$ and 
${\cal {C}}$, the calculation is much complicated and tedious. Here
we just give a sketch of it. Take indices 
$a_2=3, a_1, d_1, d_2\not= 3$,
we have
\begin{eqnarray}
&&a(\lambda -\mu )a(\lambda +\mu ){\cal {A}}_{a_1d_1}(\lambda )
{\cal {C}}_{d_2}(\mu )
-(1-\delta _{a_1d_1})b(\lambda -\mu )a(\lambda +\mu )
{\cal {C}}_{d_1}(\lambda ){\cal {A}}_{a_1d_2}(\mu )
\nonumber \\
&+&\delta _{a_1d_1}\{ b(\lambda -\mu )b(\lambda +\mu )
{\cal {D}}(\lambda ){\cal {C}}_{d_2}(\mu )
-b(\lambda -\mu )R_{21}(\lambda +\mu )_{c_1a_1}^{a_1c_1}
{\cal {C}}_{c_1}(\lambda ){\cal {A}}_{c_1d_2}(\mu )\}
\nonumber \\
&=&-{\cal {T}}(\mu )_3^3R(\lambda +\mu )_{a_13}^{3a_1}
{\cal {T}}(\lambda )_3^{c_1}R_{21}(\lambda -\mu )_{c_1a_1}^{d_1d_2}
\nonumber \\
&+&{\cal {T}}(\mu )_3^{b_2}R(\lambda +\mu )_{a_1b_2}^{b_1c_2}
{\cal {T}}(\lambda )_{b_1}^{c_1}R_{21}(\lambda -\mu )_{c_1c_2}^{d_1d_2}.
\end{eqnarray}
Subsitute the transformation (61) into this relation and 
consider it for three cases:
Case I: $a_1\not =d_1, d_1=d_2 or d_1\not= d_2;$ 
Case II: $a_1=d_1=d_2; $
Case III: $a_1=d_1\not =d_2$. The results of 
the above relation can be calculated out. However, it 
is still too complicated to 
be handled for the algebraic Bethe ansatz method. Fortunately,
we can summarize all of those results to a much concise relation:
\begin{eqnarray}
&&\tilde {\cal {A}}_{a_1d_1}(\lambda ){\cal {C}}_{d_2}(\mu )
\nonumber \\
&=&\frac {(\lambda -\mu +i)(\lambda +\mu +2i)}
{(\lambda -\mu )(\lambda +\mu +i)}
r_{12}(\lambda +\mu +i)_{a_1c_2}^{c_1b_2}
r_{21}(\lambda -\mu )_{b_1b_2}^{d_1d_2}{\cal {C}}_{c_2}(\mu )
\tilde {\cal {A}}_{c_1b_1}(\lambda )
\nonumber \\
&+&\frac {2i(\lambda +i)}{(\lambda -\mu )(2\lambda +i)}
r(2\lambda +i)_{a_1b_1}^{b_2d_1}
{\cal {C}}_{b_1}(\lambda )\tilde {\cal {A}}_{b_2d_2}(\mu )
\nonumber \\
&-&\frac {4i\mu (\lambda +i)}{(2\lambda +i)(2\mu +i)(\lambda +\mu +2i)}
r(2\lambda +i)_{a_1b_2}^{d_2d_1}{\cal {C}}_{b_2}(\lambda )
{\cal {D}}(\mu ).
\end{eqnarray}
Where the elements of the r-matrix are defined as the elements of the
original R matrix for the case all of its indices just take values 1,2.
\begin{eqnarray}
r(\lambda )_{ac}^{bd}=a(\lambda )\delta _{ab}\delta _{cd}
-b(\lambda )\delta _{ac}\delta _{bd} =a(\lambda )I+b(\lambda )\Pi ^{(1)},
\end{eqnarray}
where $\Pi ^{(1)}$ is the $4\times 4$ permutation matrix corresponding
to the grading $\epsilon _1=\epsilon _2=1$.
We can write the r-matrix as:
\begin{eqnarray}
r(\lambda )=\left( \begin{array}{cccc}
a(\lambda )-b(\lambda )&&&\\
&a(\lambda )&-b(\lambda )&\\
&-b(\lambda )&a(\lambda )&\\
&&&a(\lambda )-b(\lambda )\end{array}\right)
\end{eqnarray}

Similar as the periodic boundary condition case, we construct a set 
of the eigenvectors of the transfer matrix with reflecting boundary
conditions as: 
\begin{eqnarray}
{\cal {C}}_{d_1}(\mu _1)
{\cal {C}}_{d_2}(\mu _2)\cdots {\cal {C}}_{d_n}(\mu _n)|0>F^{d_1\cdots d_n}.
\end{eqnarray}
Here $F^{d_1\cdots d_n}$ is a function of the spectral parameters $\mu _j$.
This technique is standard for the algebraic Bethe ansatz method.
Acting the transfer matrix on this eigenvectors, we should find the
eigenvalues $\Lambda (\lambda )$
of the transfer matrix $t(\lambda )$
and a set of Bethe ansatz equations. Act first ${\cal {D}}$ on the 
eigenvector defined above, use next the commutation relation (67), 
consider the value of ${\cal {D}}$ acting on the vacuum state (58), 
we have
\begin{eqnarray}
&&{\cal {D}}(\lambda )
{\cal {C}}_{d_1}(\mu _1)
{\cal {C}}_{d_2}(\mu _2)\cdots {\cal {C}}_{d_n}(\mu _n)|0>F^{d_1\cdots d_n}
\nonumber \\
&=&U_3(\lambda )\prod _{i=1}^n
\frac {(\lambda +\mu _i)(\lambda -\mu _i-i)}
{(\lambda +\mu _i+i)(\lambda -\mu _i)}
{\cal {C}}_{d_1}(\mu _1)
{\cal {C}}_{d_2}(\mu _2)\cdots {\cal {C}}_{d_n}(\mu _n)|0>F^{d_1\cdots d_n}
+u.t.,
\end{eqnarray}
where we omited the unwanted terms $u.t.$. 

Then we act $\tilde {\cal {A}}_{aa}(\lambda )$ on the assumed eigenvector,
using repeatedly the commutation relations (69), we have
\begin{eqnarray}
&&\tilde {\cal {A}}_{aa}(\lambda )
{\cal {C}}_{d_1}(\mu _1)
{\cal {C}}_{d_2}(\mu _2)\cdots {\cal {C}}_{d_n}(\mu _n)|0>F^{d_1\cdots d_n}
\nonumber \\
&=&\prod _{i=1}^n
\frac {(\lambda -\mu _i+i)(\lambda +\mu _i+2i)}
{(\lambda -\mu _i)(\lambda +\mu _i+i)}
r_{12}(\lambda +\mu _1+i)_{ac_1}^{a_1e_1}
r_{21}(\lambda -\mu _1)_{b_1e_1}^{ad_1}
\nonumber \\
&&r_{12}(\lambda +\mu _2+i)_{a_1c_2}^{a_2e_2}
r_{21}(\lambda -\mu _2)_{b_2e_2}^{b_1d_2}
\cdots 
r_{12}(\lambda +\mu _n+i)_{a_{n-1}c_n}^{a_ne_n}
r_{21}(\lambda -\mu _n)_{b_ne_n}^{b_{n-1}d_n}
\nonumber \\
&&\times {\cal {C}}_{c_1}(\mu _1)\cdots 
{\cal {C}}_{c_n}(\mu _n)\tilde {\cal {A}}_{a_nb_n}(\lambda )|0>
F^{d_1\cdots d_n}
+u.t.. 
\end{eqnarray}
Summarize relations (50,58), we know 
\begin{eqnarray}
{\cal {A}}_{a_nb_n}(\lambda )|0>=\delta _{a_nb_n}W_{a_n}(\lambda )
a^{2L}(\lambda )|0>.
\end{eqnarray}
We can rewite the transfer matrix as:  
\begin{eqnarray}
t(\lambda )&=&-W_1^+(\lambda )\tilde {\cal {A}}_{11}(\lambda )
-W_2^+(\lambda )\tilde {\cal {A}}_{22}(\lambda )+U_3^+(\lambda ) 
{\cal {D}}(\lambda )
\nonumber \\
&=&-W_a^+(\lambda ) 
\tilde {\cal {A}}_{aa}(\lambda ).
+U_3^+(\lambda ) 
{\cal {D}}(\lambda )
\end{eqnarray}
Thus the eigenvalue of the transfer matrix with reflecting boundary
condition is written as:
\begin{eqnarray}
&&t(\lambda )
{\cal {C}}_{d_1}(\mu _1)
{\cal {C}}_{d_2}(\mu _2)\cdots {\cal {C}}_{d_n}(\mu _n)|0>F^{d_1\cdots d_n}
\nonumber \\
&=&U_3^+(\lambda )U_3(\lambda )\prod _{i=1}^n
\frac {(\lambda +\mu _i)(\lambda -\mu _i-i)}
{(\lambda +\mu _i+i)(\lambda -\mu _i)}
{\cal {C}}_{d_1}(\mu _1)
\cdots {\cal {C}}_{d_n}(\mu _n)|0>F^{d_1\cdots d_n}
\nonumber \\
&+&a^{2L}(\lambda )
\prod _{i=1}^n
\frac {(\lambda -\mu _i+i)(\lambda +\mu _i+2i)}
{(\lambda -\mu _i)(\lambda +\mu _i+i)}
{\cal {C}}_{c_1}(\mu _1)
\cdots {\cal {C}}_{c_n}(\mu _n)|0>
t^{(1)}(\lambda )^{c_1\cdots c_n}_{d_1\cdots d_n}
F^{d_1\cdots d_n}
\nonumber \\
&+&u.t.,
\end{eqnarray}
where $t^{(1)}(\lambda )$ is the so called nested transfer matrix, and
with the help of the relation (74), it can be defined as:
\begin{eqnarray}
t^{(1)}(\lambda )^{c_1\cdots c_n}_{d_1\cdots d_n}
&=&-W_a^+(\lambda )
\left\{ r(\lambda +\mu _1+i)_{ac_1}^{a_1e_1}
r(\lambda +\mu _2+i)_{a_1c_2}^{a_2e_2}\cdots
r(\lambda +\mu _1+i)_{a_{n-1}c_n}^{a_ne_n}\right\}
\nonumber \\
&&\delta _{a_nb_n}W_{a_n}(\lambda )
\left\{ r_{21}(\lambda -\mu _n)_{b_ne_n}^{b_{n-1}d_n}
\cdots
r_{21}(\lambda -\mu _2)_{b_2e_2}^{b_1d_2}
r_{21}(\lambda -\mu _1)_{b_1e_1}^{ad_1}\right\} .
\end{eqnarray}
We find that this nested transfer matrix can be defined as
a transfer matrix with reflecting boundary conditions corresponding
to the anisotropic case
\begin{eqnarray}
t^{(1)}(\lambda )=str{K^{(1)}}^+(\tilde {\lambda })
T^{(1)}(\tilde {\lambda }, \{ \tilde {\mu }_i\} )
K^{(1)}(\tilde {\lambda })
{T^{(1)}}^{-1}(-\tilde {\lambda }, 
\{ \tilde {\mu }_i\} )
\end{eqnarray}
with the grading $\epsilon _1=\epsilon _2=1$, where we denote
$\tilde {\lambda }=\lambda +{i\over 2}, \tilde {\xi }=\xi +{i\over 2},
\tilde {\xi }^+=\xi ^+-{i\over 2}$, we will also denote 
$\tilde {\mu }=\mu +{i\over 2}$ later.
Explicitly we have 
${K^{(1)}}^+_I(\tilde {\lambda })=id.$ up to a whole factor 
$\tilde {\xi }^++i-\tilde {\lambda }$, and 
\begin{eqnarray}
{K^{(1)}}^+_{II}(\tilde {\lambda })=\left(
\begin{array}{cc}W_1^+(\tilde {\lambda })&\\ 
&W_2^+(\tilde {\lambda })\end{array}\right)
=\left(
\begin{array}{cc}
\tilde {\xi }^++i-\tilde {\lambda }&\\
&\tilde {\xi }^+-i+\tilde {\lambda }\end{array}
\right) 
\end{eqnarray}
corresponding to $K_I^+$ and $K_{II}^+$ respectively. 
We also have 
$K^{(1)}_I(\tilde {\lambda })=id.$ up to a whole factor  
$\frac {(2\tilde {\lambda }-i)(\tilde {\lambda }+\tilde {\xi })}
{2\tilde {\lambda }}$, and
\begin{eqnarray}
K^{(1)}_{II}(\tilde {\lambda })=
\left( \begin{array}{cc}W_1(\tilde {\lambda })&\\ 
&W_2(\tilde {\lambda })\end{array}\right)
=\frac {(2\tilde {\lambda }-i)}{2\tilde {\lambda }}
\left(
\begin{array}{cc}
\tilde {\xi }+\tilde {\lambda }&\\
&\tilde {\xi }-\tilde {\lambda }\end{array}
\right) 
\end{eqnarray}
corresponding to $K_I$ and $K_{II}$.
The row-to-row monodromy matrix 
$T^{(1)}(\tilde {\lambda }, \{ \tilde {\mu }_i\} )$ (corresponding
to the periodic boundary condition) is defined as:
\begin{eqnarray}
T^{(1)}_{aa_n}(\tilde {\lambda }, 
\{ \tilde {\mu }_i\} )_{c_1\cdots c_n}^{e_1\cdots e_n}
&=&
r(\tilde {\lambda }+\tilde {\mu }_1)_{ac_1}^{a_1e_1}
r(\tilde {\lambda }+\tilde {\mu }_2)_{a_1c_2}^{a_2e_2}\cdots
r(\tilde {\lambda }+\tilde {\mu }_1)_{a_{n-1}c_n}^{a_ne_n}
\nonumber \\
&=&L_1^{(1)}(\tilde {\lambda }+\tilde {\mu }_1)
L_2^{(1)}(\tilde {\lambda }+\tilde {\mu }_2)\cdots
L_n^{(1)}(\tilde {\lambda }+\tilde {\mu }_1).
\end{eqnarray}
The L-operator takes the form
\begin{eqnarray}
L^{(1)}_k(\tilde {\lambda })=
\left(
\begin{array}{cc}a(\tilde {\lambda })-b(\tilde {\lambda })e_k^{11}
&-b(\tilde {\lambda })e_k^{21}\\
-b(\tilde {\lambda })e_k^{12}&
a(\tilde {\lambda })-b(\tilde {\lambda })e_k^{22}\end{array}
\right) .
\end{eqnarray}
And we also have
\begin{eqnarray}
{T^{(1)}}^{-1}(-\tilde {\lambda }, 
\{ \tilde {\mu }_i\} )
&=&
r_{21}(\tilde {\lambda }-\tilde {\mu }_n)_{b_ne_n}^{b_{n-1}d_n}
\cdots
r_{21}(\tilde {\lambda }-\tilde {\mu }_2)_{b_2e_2}^{b_1d_2}
r_{21}(\tilde {\lambda }-\tilde {\mu }_1)_{b_1e_1}^{ad_1}
\nonumber \\
&=&{L_n^{(1)}}^{-1}(-\tilde {\lambda }+\tilde {\mu }_n )
\cdots 
{L_2^{(1)}}^{-1}(-\tilde {\lambda }+\tilde {\mu }_2 )
{L_1^{(1)}}^{-1}(-\tilde {\lambda }+\tilde {\mu }_1 ),
\end{eqnarray}
where we have used the 
unitarity relation of the r-matrix 
$r_{12}(\lambda )r_{21}(-\lambda )=1$.

In this section, We show that the problem to find the eigenvalue 
of the original transfer matrix $t(\lambda )$
become the problem to find the eignevalue of the nested transfer
matrix $t^{(1)}(\lambda )$. In relation (77), one can see that
besides the wanted term which is dedicated to the eigenvalue, we also
have the unwanted terms which must be canceled so that the assumed
eigenvector is indeed the eigenvector of the transfer matrix. With
the help of the symmetry property (66) of the assumed eigenvector (72),
use the standard algebraic Bethe ansatz method, we find if 
$\mu _1, \cdots, \mu _n$ satisfy the following Bethe ansatz equations,
the unwanted terms will vanish.
\begin{eqnarray}
a^{-2L}(\mu _j)U_3(\mu _j)U_3^+(\mu _j)\frac {\mu _j}{\mu _j+i}
\prod _{i=1, i\not =j}^n
\frac {(\mu _j+\mu _i)(\mu _j-\mu _i-i)}
{(\mu _j-\mu _i+i)(\mu _j+\mu _i+2i)}
=\Lambda ^{(1)}(\mu _j), 
\nonumber \\
j=1, 2,\cdots ,n.
\end{eqnarray}
Here we have used the notation $\Lambda ^{(1)}$ to denote the eigenvalue
of the nested transfer matrix $t^{(1)}(\lambda )$.

Thus what we should do next is to find the eigenvalue of the nested
transfer matrix $t^{(1)}$.

\subsection{The nested algebraic Bethe ansatz method}
We hope that the eigenvalue of the nested transfer matrix can be solved
similarly as that of the original transfer matrix. It seems that there is
a logic error to call $t^{(1)}$ the transfer matrix, because that we
did not show $t^{(1)}$ commute with each other for different spectral
parameters. On the other hand, $K^{(1)}$ and ${K^{(1)}}^+$ have already
been defined explicitly in the above section, they are not obtained
from, for example, reflection equation and the dual reflection equation.
In this section, we will shown that all of those
problems can be solved in the framework of the reflecting boundary
condition case.

We know that the following graded Yang-Baxter relation with 
grading $\epsilon _1=\epsilon _2=1$ is satisfied:
\begin{eqnarray}
r(\lambda -\mu )L_1^{(1)}(\lambda )L_2^{(1)}(\mu )
=L_2^{(1)}(\mu )L_1^{(1)}(\lambda )r(\lambda -\mu )
\end{eqnarray}
So, we also have the Yang-Baxter relation for the row-to-row monodromy
matrix 
\begin{eqnarray}
r(\lambda -\mu )
T^{(1)}_1(\lambda , \{ \mu _i\} )
T^{(1)}_2(\mu , \{ \mu _i\} )
=T^{(1)}_2(\mu , \{ \mu _i\} )
T^{(1)}_1(\lambda , \{ \mu _i\} )
r(\lambda -\mu ).
\end{eqnarray}
We propose the reflection equation take the form
\begin{eqnarray}
&&r_{12}(\lambda -\mu )K_1^{(1)}(\lambda )r_{21}(\lambda +\mu )
K_2^{(1)}(\mu )
\nonumber \\
&=&K_2^{(1)}(\mu )r_{12}(\lambda +\mu )K_1^{(1)}(\lambda )
r_{21}(\lambda -\mu ).
\end{eqnarray}
Solving this reflection equation directly, we find 
\begin{eqnarray}
K^{(1)}(\lambda )=
\left(
\begin{array}{cc}
\xi +\lambda &\\
&\xi -\lambda \end{array}
\right) 
\end{eqnarray}
is a solution of the reflection equation, it is easy to show that
$K^{(1)}=id.$ is also a solution. These solutions of the reflection
equation are consistent with the results defined in the above subsetion.
So, we can show that the nested double-row monodromy matrix 
\begin{eqnarray}
{\cal {T}}^{(1)}(\lambda ,\{ \mu _i\} )\equiv 
T^{(1)}(\lambda ,\{ \mu _i\} )K^{(1)}(\lambda )
{T^{(1)}}^{-1}(-\lambda ,\{ \mu _i\} ) 
\end{eqnarray}
also satisfy the the reflection equation,
\begin{eqnarray}
&&r_{12}(\lambda -\mu ){\cal {T}}^{(1)}_1(\lambda ,\{ \mu _i\} )
r_{21}(\lambda +\mu )
{\cal {T}}^{(1)}_2(\mu ,\{ \mu _i\} )
\nonumber \\
&=&{\cal {T}}^{(1)}_2(\nu ,\{ \mu _i\} )
r_{12}(\lambda +\mu )
{\cal {T}}^{(1)}_1(\lambda ,\{ \mu _i\} )r_{21}(\lambda -\mu ).
\end{eqnarray}

One can prove that the r-matrix satisfy a unitarity relation
\begin{eqnarray}
r_{12}^{st_1}(\lambda )r_{21}^{st_1}(2i-\lambda )=
a(\lambda )a(2i-\lambda )\cdot id.
\end{eqnarray}
According to this relation and the unitarity relation of
the r-matrix $r_{12}(\lambda )r_{21}(-\lambda )=1\cdot id.$, 
we propose the following
dual reflection equation 
\begin{eqnarray}
&&r_{12}(\mu -\lambda ){K_1^{(1)}}^+(\lambda )
r_{21}(\lambda +\mu +2i)
{K_2^{(1)}}^+(\mu )
\nonumber \\
&=&{K_2^{(1)}}^+(\mu )r_{12}(\lambda +\mu +2i)
{K_1^{(1)}}^+(\lambda )
r_{21}(\mu -\lambda ).
\end{eqnarray}
We can also find that the solution of the dual reflection equation is
consistent with the ${K^{(1)}}^+$ results (80)
presented in the above subsetion.
Similar procedure as that presented in section III
can also be applied now, we find that
the defined nested transfer matrix indeed consitute a commuting family for
different spectral parameters.

Now, let us use again the algebraic Bethe ansatz method to obtain
the eigenvalue $\Lambda ^{(1)}(\lambda )$ of the nested transfer matrix
$t^{(1)}(\lambda )$. We write the nested double-row monodromy matrix as:
\begin{eqnarray}
{\cal {T}}^{(1)}(\lambda ,\{ \mu _i\} )
&=&\left( \begin{array}{cc}
{\cal {A}}^{(1)}(\lambda ) &{\cal {B}}^{(1)}(\lambda ) \\
{\cal {C}}^{(1)}(\lambda ) &{\cal {D}}^{(1)}(\lambda ) \end{array}
\right)  
\nonumber \\
&=&T^{(1)}(\lambda ,\{ \mu _i\} )K^{(1)}(\lambda )
{T^{(1)}}^{-1}(-\lambda ,\{ \mu _i\} ) 
\nonumber \\
&=&\left( \begin{array}{cc}
A^{(1)}(\lambda ) &B^{(1)}(\lambda ) \\
C^{(1)}(\lambda ) &D^{(1)}(\lambda ) \end{array}
\right) \left( 
\begin{array}{cc}k^{(1)}_1(\lambda ) &\\
&k_2^{(1)}(\lambda )\end{array}\right)
\left( \begin{array}{cc}
{\bar {A}}^{(1)}(-\lambda ) &\bar {B}^{(1)}(-\lambda ) \\
\bar {C}^{(1)}(-\lambda ) &\bar {D}^{(1)}(-\lambda ) \end{array}
\right) \nonumber \\  
\end{eqnarray}
For convenience, we introduce again a transformation 
\begin{eqnarray}
{\cal {A}}^{(1)}(\lambda )=\tilde {\cal {A}}^{(1)}(\lambda )
-\frac {i}{2\lambda -i}{\cal {D}}^{(1)}(\lambda ).
\end{eqnarray}
Because that the nested double-row monodromy matrix satisfy
the reflection equation (91), we can find the following commutation
relations:
\begin{eqnarray}
{\cal {D}}^{(1)}(\lambda )
{\cal {C}}^{(1)}(\mu )
&=&\frac {(\lambda -\mu +i)(\lambda +\mu )}
{(\lambda -\mu )(\lambda +\mu -i)}
{\cal {C}}^{(1)}(\mu )
{\cal {D}}^{(1)}(\lambda )
\nonumber \\
&-&\frac {2i\mu }{(\lambda -\mu )(2\mu -i)}
{\cal {C}}^{(1)}(\lambda )
{\cal {D}}^{(1)}(\mu )
+\frac {i}{\lambda +\mu -i}
{\cal {C}}^{(1)}(\lambda )
\tilde {\cal {A}}^{(1)}(\mu ),\\
\tilde {\cal {A}}^{(1)}(\lambda )
{\cal {C}}^{(1)}(\mu )
&=&\frac {(\lambda -\mu -i)(\lambda +\mu -2i)}
{(\lambda -\mu )(\lambda +\mu -i)}
{\cal {C}}^{(1)}(\mu )
\tilde {\cal {A}}^{(1)}(\lambda )
\nonumber \\
&+&\frac {2i(\lambda -i)}{(\lambda -\mu )(2\lambda -i)}
{\cal {C}}^{(1)}(\lambda )
\tilde {\cal {A}}^{(1)}(\mu )
-\frac {4i\mu (\lambda -i)}{(\lambda +\mu -i)(2\lambda -i)(2\mu -i)},
\\
{\cal {C}}^{(1)}(\lambda )
{\cal {C}}^{(1)}(\mu )
&=&
{\cal {C}}^{(1)}(\mu )
{\cal {C}}^{(1)}(\lambda  ).
\end{eqnarray}
As the reference states, for the nesting we pick
\begin{eqnarray}
|0>^{(1)}_k=\left( \begin{array}{c}0\\1\end{array}\right),
|0>^{(1)}=\otimes _{k=1}^n|0>_k^{(1)}.
\end{eqnarray}
With the help of the definition (82, 84), 
we know the results of the nested monodromy matrix and the inverse
of the monodromy matrix acting on the reference state
\begin{eqnarray}
T^{(1)}(\lambda ,\{ \mu _i\} )|0>^{(1)}
&=&
\left( \begin{array}{cc}
A^{(1)}(\tilde {\lambda }) &B^{(1)}(\tilde {\lambda }) \\
C^{(1)}(\tilde {\lambda }) &D^{(1)}(\tilde {\lambda }) \end{array}
\right)|0>^{(1)}
\nonumber \\
&=& 
\left( \begin{array}{cc}
\prod _{i=1}^na(\tilde {\lambda }+\tilde {\mu }_i) &0 \\
C^{(1)}(\tilde {\lambda }) &
\prod _{i=1}^n[a(\tilde {\lambda }+\tilde {\mu }_i)-
b(\tilde {\lambda }+\tilde {\mu })]\end{array}
\right)|0>^{(1)}, 
\nonumber \\
\\
{T^{(1)}}^{-1}(-\lambda ,\{ \mu _i\} )|0>^{(1)} 
&=&\left( 
\begin{array}{cc}
\bar {A}^{(1)}(\tilde {\lambda }) &\bar {B}^{(1)}(\tilde {\lambda }) \\
\bar {C}^{(1)}(\tilde {\lambda }) &
\bar {D}^{(1)}(\tilde {\lambda }) \end{array}
\right)|0>^{(1)}
\nonumber \\
&=& 
\left(\begin{array}{cc}
\prod _{i=1}^na(\tilde {\lambda }-\tilde {\mu }_i) &0 \\
C^{(1)}(\tilde {\lambda }) &
\prod _{i=1}^n[a(\tilde {\lambda }-\tilde {\mu }_i)-
b(\tilde {\lambda }-\tilde {\mu })]\end{array}
\right)|0>^{(1)} .
\nonumber \\
\end{eqnarray}
Substituting those relation into relation (94), 
we can obtain 
the results of the nested double-row monodromy matrix acting
on the nested vacuum state $|0>^{(1)}$, 
\begin{eqnarray}
{\cal {B}}^{(1)}(\tilde {\lambda })|0>^{(1)}=0,~~~ 
{\cal {C}}^{(1)}(\tilde {\lambda })|0>^{(1)}\not=0, 
\end{eqnarray}
\begin{eqnarray}
{\cal {D}}^{(1)}(\tilde {\lambda })|0>^{(1)}= 
U_2(\tilde {\lambda })\prod _{i=1}^n
\left\{ [a(\tilde {\lambda }+\tilde {\mu }_i)-
b(\tilde {\lambda }+\tilde {\mu }_i)]
[a(\tilde {\lambda }-\tilde {\mu }_i)-
b(\tilde {\lambda }-\tilde {\mu }_i)]\right\} |0>^{(1)}.
\nonumber \\
\end{eqnarray}
Here we use the notation $U_2=k^{(1)}_2$, 
$U_2(\tilde {\lambda })=\frac {(2\tilde {\lambda }-i)(\tilde {\lambda }
+\tilde {\xi })}{2\tilde {\lambda }}$ for $K_I$ case, 
$U_2(\tilde {\lambda })=\frac {(2\tilde {\lambda }-i)
(\tilde {\xi }-\tilde {\lambda })}{2\tilde {\lambda }}$ for $K_{II}$ case. 

The result of element ${\tilde {A}}^{(1)}$ is not so direct. We know
\begin{eqnarray}
{\cal {A}}^{(1)}(\tilde {\lambda })|0>^{(1)}
=k_1^{(1)}(\tilde {\lambda })A^{(1)}(\tilde {\lambda })
\bar {A}^{(1)}(-\tilde {\lambda })|0>^{(1)}
+k_2^{(1)}(\tilde {\lambda })
B^{(1)}(\tilde {\lambda })
\bar {C}^{(1)}(-\tilde {\lambda })|0>^{(1)}.
\end{eqnarray}
Using the Yang-Baxter relation (87), 
the above relation becomes
\begin{eqnarray}
{\cal {A}}^{(1)}(\tilde {\lambda })|0>^{(1)}
&=&k_1^{(1)}(\tilde {\lambda })A^{(1)}(\tilde {\lambda })
\bar {A}^{(1)}(-\tilde {\lambda })|0>^{(1)}
\nonumber \\
&&+k_2^{(1)}(\tilde {\lambda })
\frac {b(2\tilde {\lambda })}{a(2\tilde {\lambda })-
b(2\tilde {\lambda })}
[A^{(1)}(\tilde {\lambda })
\bar {A}^{(1)}(-\tilde {\lambda })
-\bar {D}^{(1)}(-\tilde {\lambda })
D^{(1)}(\tilde {\lambda })]|0>^{(1)}
\nonumber \\
&=&[k_1^{(1)}(\tilde {\lambda })
+k_2^{(1)}(\tilde {\lambda })
\frac {b(2\tilde {\lambda })}{a(2\tilde {\lambda })-
b(2\tilde {\lambda })}]
\prod _{i=1}^n[a(\tilde {\lambda }+\tilde {\mu }_i)
a(\tilde {\lambda }-\tilde {\mu }_i)]|0>^{(1)}
\nonumber \\
&&-\frac {i}{2\tilde {\lambda }-i}
{\cal {D}}^{(1)}(\tilde {\lambda })|0>^{(1)} .
\end{eqnarray}
With the help of the transformation (94), we find
\begin{eqnarray}
\tilde {\cal {A}}^{(1)}(\tilde {\lambda })|0>^{(1)}
=U_1(\tilde {\lambda })
\prod _{i=1}^n[a(\tilde {\lambda }+\tilde {\mu }_i)
a(\tilde {\lambda }-\tilde {\mu }_i)]|0>^{(1)},
\end{eqnarray}
where we denote $U_1(\tilde {\lambda })=k_1^{(1)}(\tilde {\lambda })
+k_2^{(1)}(\tilde {\lambda })
\frac {b(2\tilde {\lambda })}{a(2\tilde {\lambda })-
b(2\tilde {\lambda })}$, and $U_1$ take the following form
explicitly

For $K_I(\lambda )$:
\begin{eqnarray}
U_1(\tilde {\lambda })=\tilde {\lambda }+\tilde {\xi },
\end{eqnarray}
\indent For $K_{II}(\lambda )$:
\begin{eqnarray}
U_1(\tilde {\lambda })=\tilde {\lambda }+\tilde {\xi }-i.
\end{eqnarray}
The nested transfer matrix takes the form 
\begin{eqnarray}
t^{(1)}(\tilde {\lambda })&=&strK^{(1)}(\tilde {\lambda })
{\cal {T}}^{(1)}(\tilde {\lambda })
\nonumber \\
&=&-{k_1^{(1)}}^+(\tilde {\lambda }){\cal {A}}^{(1)}(\tilde {\lambda })
-{k_2^{(1)}}^+(\tilde {\lambda }){\cal {D}}^{(1)}(\tilde {\lambda })
\nonumber \\
&=&-U_1^+(\tilde {\lambda })
\tilde {\cal {A}}^{(1)}(\tilde {\lambda })
-U_2^+(\tilde {\lambda })
{\cal {D}}^{(1)}(\tilde {\lambda }),
\end{eqnarray}
where we denote $U_1^+={k_1^{(1)}}^+$, 
$U_2^+(\lambda )={k_2^{(1)}}^+(\lambda )-
\frac {i}{2\lambda -i}{k_1^{(1)}}^+(\lambda )$, that means:

For $K^+_I$ case:
\begin{eqnarray}
U_1^+(\tilde {\lambda })&=&\tilde {\xi }^++i-\tilde {\lambda },
\\
U_2^+(\tilde {\lambda })&=&
\frac {2(\tilde {\lambda }-i)
(\tilde {\xi }^++i-\tilde {\lambda })
}{2\tilde {\lambda }-i},
\end{eqnarray}
\indent For $K^+_{II}$ case:
\begin{eqnarray}
U_1^+(\tilde {\lambda })&=&\tilde {\xi }^++i-\tilde {\lambda },
\nonumber \\
U_2^+(\tilde {\lambda })&=&
\frac {2(\tilde {\lambda }+\tilde {\xi }^+)(\tilde {\lambda }-i)}
{2\tilde {\lambda }-i}.
\end{eqnarray}

Using the standard algebraic Bethe ansatz method, assume that the
eigenvector of the nested transfer matrix constucted as
${\cal {C}}(\tilde {\mu }^{(1)}_1)
{\cal {C}}(\tilde {\mu }^{(1)}_2)\cdots 
{\cal {C}}(\tilde {\mu }^{(1)}_m)|0>^{(1)}$. Acting the nested transfer
matrix on this eigenvector, using repeatedly the commutation
relations (96,97), we have the eigenvalue
\begin{eqnarray}
\Lambda ^{(1)}(\tilde {\lambda })
&=&-U_1^+(\tilde {\lambda })
U_1(\tilde {\lambda })
\prod _{i=1}^n[a(\tilde {\lambda }+\tilde {\mu }_i)
a(\tilde {\lambda }-\tilde {\mu }_i)]
\prod _{l=1}^m \left\{
\frac {(\tilde {\lambda }-\tilde {\mu }^{(1)}_l+i)
(\tilde {\lambda }+\tilde {\mu }^{(1)}_l)}
{(\tilde {\lambda }-\tilde {\mu }^{(1)}_l)
(\tilde {\lambda }+\tilde {\mu }^{(1)}_l-i)}\right\} 
\nonumber\\ 
&&-U_2^+(\tilde {\lambda })
U_2(\tilde {\lambda })\prod _{i=1}^n
\left\{ [a(\tilde {\lambda }+\tilde {\mu }_i)-
b(\tilde {\lambda }+\tilde {\mu }_i)]
[a(\tilde {\lambda }-\tilde {\mu }_i)-
b(\tilde {\lambda }-\tilde {\mu }_i)]\right\} 
\nonumber \\
&&\prod _{l=1}^m
\left\{ \frac {(\tilde {\lambda }-\tilde {\mu }^{(1)}_l-i)
(\tilde {\lambda }+\tilde {\mu }^{(1)}_l-2i)}
{(\tilde {\lambda }-\tilde {\mu }^{(1)}_l)
(\tilde {\lambda }+\tilde {\mu }^{(1)}_l-i)}\right\} ,
\end{eqnarray}
where $\tilde {\mu }^{(1)}_1, \cdots ,\tilde {\mu }^{(1)}_m$
should satisfy the following Bethe ansatz equations
\begin{eqnarray}
&&\frac {U_1^+(\tilde {\mu }_j^{(1)})
U_1(\tilde {\mu }_j^{(1)})}
{U_2^+(\tilde {\mu }_j^{(1)})
U_2(\tilde {\mu }_j^{(1)})}
\frac {\tilde {\mu }_j^{(1)}}{(\tilde {\mu }_j^{(1)}-i)}
\prod _{i=1}^n
\frac {(\tilde {\mu }_j^{(1)}+\tilde {\mu }_i)
(\tilde {\mu }_j^{(1)}-\tilde {\mu }_i)}
{(\tilde {\mu }_j^{(1)}+\tilde {\mu }_i-i)
(\tilde {\mu }_j^{(1)}-\tilde {\mu }_i-i)} 
\nonumber \\
&=&
\prod _{l=1, \not=j}^m
\frac {(\tilde {\mu }_j^{(1)}-\tilde {\mu }^{(1)}_l-i)
(\tilde {\mu }_j^{(1)}+\tilde {\mu }^{(1)}_l-2i)}
{\tilde {\mu }_j^{(1)}-\tilde {\mu }^{(1)}_l+i)
(\tilde {\mu }_j^{(1)}+\tilde {\mu }^{(1)}_l)}, ~~~j=1, \cdots , m.
\end{eqnarray}

Thus, the eigenvalue of the transfer matrix $t(\lambda )$
with reflecting boundary condition (36) is obtained as:
\begin{eqnarray}
\Lambda (\lambda )
&=&U_3^+(\lambda )U_3(\lambda )\prod _{i=1}^n
\frac {(\lambda +\mu _i)(\lambda -\mu _i-i)}
{(\lambda +\mu _i+i)(\lambda -\mu _i)}
\nonumber \\
&+&a^{2L}(\lambda )
\prod _{i=1}^n
\frac {(\lambda -\mu _i+i)(\lambda +\mu _i+2i)}
{(\lambda -\mu _i)(\lambda +\mu _i+i)}
\Lambda ^{(1)}(\tilde {\lambda }).
\end{eqnarray}
Here for convenience, we give summary of the values $U$ and $U^+$. 

Case I:
\begin{eqnarray}
U_1^+(\tilde {\lambda })&=&\tilde {\xi }^++i-\tilde {\lambda },
\nonumber \\
U_2^+(\tilde {\lambda })&=&
\frac {2(\tilde {\lambda }-i)
(\tilde {\xi }^++i-\tilde {\lambda })
}{2\tilde {\lambda }-i},
\nonumber \\
U_3^+(\lambda )&=&\frac {(2\lambda -i)(\xi ^++\lambda +i)}{2\lambda +i},
\end{eqnarray}

Case II:
\begin{eqnarray}
U_1^+(\tilde {\lambda })&=&\tilde {\xi }^++i-\tilde {\lambda },
\nonumber \\
U_2^+(\tilde {\lambda })&=&
\frac {2(\tilde {\lambda }+\tilde {\xi }^+)(\tilde {\lambda }-i)}
{2\tilde {\lambda }-i}.
\nonumber \\
U_3^+(\lambda )&=&\frac {(2\lambda -i)(\xi ^++\lambda )}{2\lambda +i}.
\end{eqnarray}

Case I:
\begin{eqnarray}
U_1(\tilde {\lambda })&=&\tilde {\lambda }+\tilde {\xi },
\nonumber \\
U_2(\tilde {\lambda })&=&
\frac {(2\tilde {\lambda }-i)(\tilde {\lambda }
+\tilde {\xi })}{2\tilde {\lambda }},
\nonumber \\
U_3(\lambda )&=&(\xi -\lambda )
\end{eqnarray}

Case II:
\begin{eqnarray}
U_1(\tilde {\lambda })&=&\tilde {\lambda }+\tilde {\xi }-i.
\nonumber \\
U_2(\tilde {\lambda })&=&\frac {(2\tilde {\lambda }-i)
(\tilde {\xi }-\tilde {\lambda })}{2\tilde {\lambda }},
\nonumber \\
U_3(\lambda )&=&(\xi -\lambda ).
\end{eqnarray}
We know that U and $U^+$ are independent of each other, so there
are four combinations for $\{ U, U^+\} $ such as $\{I, I\}$,
$\{I, II\}$, $\{II, I\}$ and $\{II, II\}$.

\section{Summary and discussions}
In this paper, we study the supersymmetric $t-J$ model with 
reflecting boundary conditions. We first studied the unitarity
relation and the cross-unitarity relation of the R matrix. 
According those 
relations, we proposed the reflection equation and the 
dual reflection equation for this supersymmetric $t-J$ model.
Solving the reflection equation and to dual reflection equation
we give two types solutions for them respectively. The transfer
matrix for the supersymmetric $t-J$ model is then constructed,
and we proved that the transfer matrix constitute a commuting
family. Using the nested algebraic Bethe ansatz method, we
obtain the eigenvalues of the transfer matrix. 

We discussed all of the above in the FFB grading. For the periodic
boundary conditions, the supersymmetric $t-J$ model was studied
in three different background grading.$^3$ We can also study
the FBF and BFF grading for the supersymmetric $t-J$ model with
reflecting boundary conditions. The integrability can be
proved similar as that of in this paper. We have found the
unitarity and cross-unitarity relations of the R matrix
take the same form as that
of FFB grading, we also have a same solutions for the reflection 
equation. Using similar nested algebraci Bethe ansatz method,
we can find the eigenvalues of the transfer matrix with reflecting
boundary conditions with grading FBF and BFF.

The R-matrix of the supersymmetric $t-J$ model are rational R-matrix,
there are trigonometric R-matrix corresponding to a generalized 
supersymmetric $t-J$ model. We can also study this generalized  
supersymmetric $t-J$ model with reflecting boundary conditions.

It is interest to continue study the thermodynamic limit of the
result obtained in this paper. Thus we can find some physical 
quantites such as free energy, surface free energy and
interfacial tension ect..

We can also extend the supersymmetric $t-J$ model to a more
general supersymmetric case. The R-matrix will equivalent to
the R-matrix of the Perk-Shultz model.$^{12}$

\vskip 1truecm
{\bf Acknowlegements}: This work is supported in part by the
National Natural Science Foundation of China.

\end{document}